\documentclass[final]{article}
\usepackage{anysize,subfigure}

\marginsize{1.2in}{1.2in}{1.75cm}{1.75cm}

\usepackage{amssymb}
\usepackage{graphicx}
\usepackage[printonlyused]{acronym}
\usepackage{listings}
\usepackage{url}
\usepackage[full]{complexity}
\usepackage{colortbl}
\usepackage{tikz}
\usepackage{pdflscape}
\usepackage{xcolor}

\newcommand{\T}{${\mathcal T}$}
\newcommand{\SSS}{${\mathcal S}$}
\newcommand{\CC}{${\mathcal C}$}
\input{Qcircuit.tex}
\renewcommand{\Qcircuit}[1][0em]{\xymatrix @*=<#1>}
\def\imagetop#1{\vtop{\null\hbox{#1}}}
\begin{document}

\title{\large Constant-Factor Optimization of Quantum Adders on 2D Quantum Architectures}
\author{\large Mehdi Saeedi\footnote{msaeedi@usc.edu}, Alireza Shafaei, Massoud Pedram\\\\
\large Department of Electrical Engineering, University of Southern California,\\
\large Los Angeles, CA 90089-2562\\
}

\maketitle

\normalsize
\begin{abstract}
Quantum arithmetic circuits have practical applications in various quantum algorithms. In this paper, we address quantum addition on 2-dimensional nearest-neighbor architectures based on the work presented by Choi and Van Meter (JETC 2012). To this end, we propose new circuit structures for some basic blocks in the adder, and reduce communication overhead by adding concurrency to consecutive blocks and also by parallel execution of expensive Toffoli gates. The proposed optimizations reduce total depth from $140\sqrt n+k_1$ to $92\sqrt n+k_2$ for constants $k_1,k_2$ and affect the computation fidelity considerably.
\end{abstract}

\section{Introduction} \label{sec:intro}

Quantum algorithms are often described in the quantum circuit model of computation, where for a quantum circuit with $n$ qubits, any pairs of qubits can interact. However, current advances in physical quantum technologies can only allow qubit interactions in one-, two-, or three-dimensional spaces. Restricting interactions to only linear dimension results in $O(n)$ overhead. On the other hand, working with 2D (or 3D) quantum architectures where each qubit can interact with 4 (or 6) neighboring qubits provides more flexibility.

For a given quantum circuit $C$ one can construct an interaction graph $G_{C}=(V_C,E_C)$, the nodes of which represent qubits in $C$ with edges between them when a gate in $C$ involves the related qubits. Additionally, the architecture (or fabric) of a quantum computing system can be described by a simple connected graph $G_Q = (V_Q,E_Q)$ where vertices $V_Q$ represent qubits and edges $E_Q$ represent adjacent qubit pairs that gates can be applied on \cite{Cheung07}. Accordingly, the problem of mapping a quantum circuit $C$ with arbitrary interactions between qubits onto a quantum architecture with limited interaction distance can be mapped to the problem of embedding graph $G_C$ into graph $G_Q$.

In general, the graph embedding problem is \NP-hard. However, optimal embedding methods with polynomial time complexities for several classes of graphs have been proposed \cite{DiazPS02}. In \cite{Choi:2011}, the concept of \emph{dilation} in graph embedding has been applied to find a depth lower bound for a quantum circuit after embedding. In this case, dilation is defined as the maximum distance between adjacent nodes of the graph after embedding. Working with proven properties of log-depth binary trees and considering the fact that log-depth quantum addition circuits exist, Choi and Van Meter \cite{Choi:2011} showed that the depth lower bound of the exact quantum addition circuit on a $k$-dimensional quantum architecture is $\Omega(\sqrt[k]{n})$. In \cite{Beals2012}, the authors examined the minimum overhead in depth for emulating a circuit $C$ by a circuit $C'$ subject to the constraints imposed by the interaction constraints and showed that this overhead is $O(n)$ for 1D, $O(\sqrt n)$ for 2D, $O(\log^2 n)$ or $O(\log n)$ (depending on the approach) for hypercube.

Exploring an efficient realization of a given quantum algorithm or quantum circuit for a restricted architecture has been followed by a number of researchers during the recent years. Physical implementations of the quantum Fourier transform (QFT) \cite{Takahashi:2007,Maslov07}, Shor's factorization algorithm \cite{FDH:2004,Kutin:2007,Pham2012}, quantum error correction \cite{Fowler}, and general reversible circuits \cite{ArabzadehQIP2013} for 1D/2D architectures have been explored in the past. Worst-case synthesis cost of a general/Boolean unitary matrix under the 1D restriction has been discussed in \cite{mottonen06,Shende06,SaeediQIC11,SaeediJETC10}. In \cite{SaeediQIP11,HirataQIC11,Shafaei_13} heuristic methods for converting an arbitrary quantum circuit to its equivalent circuit on 1D architectures have been proposed.

Quantum adder and its modular version have applications in different quantum algorithms including Shor's factoring algorithm. In \cite{Choi:2012}, a quantum adder with $\Theta(\sqrt n)$ depth on 2D quantum architectures was proposed which has $140 \sqrt n - 72$ depth, in terms of one- and two-qubit quantum gates. Asymptotically, the depth of the proposed adder is optimal. However, constant-factor optimization is possible and in fact desirable. Besides the effect of reducing circuit size/depth on physical realization, any additional gate in the circuit longest path can reduce circuit fidelity to some extent. Based on the analysis done in \cite{Szkopek} for fault-tolerant error correction with a concatenated 7-qubit CSS code \cite{Nielsen00}, nearest-neighbour communication overhead results in {175x} reduction in error threshold. Improving error threshold is costly and may include using a more sophisticated quantum control protocol to have gates with higher fidelities or applying a more robust error correction code. Therefore, reducing unnecessary communication overhead for a useful quantum computation is vital. Because of the effect of addition on e.g., modular multiplication and modular exponentiation circuits \cite{Pham2012,modmultqic,Markov2013}, reducing communication overhead for quantum adder by circuit optimization --- the focus of this work --- is of particular interest.

\begin{figure}
\centering
\scalebox{0.9}{
\scriptsize
\input{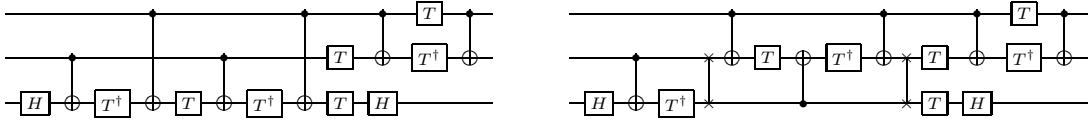}
}
\caption{Decomposition of the Toffoli gate into one-qubit and six CNOT gates \cite{Shende09} and the implementation with adjacent qubits.}
\label{fig:Toffoli}
\end{figure}

In this paper, we show how $140 \sqrt n $+const depth in \cite{Choi:2012} can be further improved to $92 \sqrt n$+const. For this purpose, we reconsider the basic blocks in the suggested quantum adder and introduce some constant-factor optimizations in communication overhead in different stages. To physically implement a given circuit, one needs to decompose all gates into primitive one- and two-qubit gates. To decompose a 3-qubit Toffoli (\T) gate, we use Clifford+T gates which are universal and have fault-tolerant (FT) implementation \cite{Nielsen00}. Figure \ref{fig:Toffoli} shows the decomposition of the Toffoli gate into one- and two-qubit gates. To consider depth, we report circuit depth in terms of single-qubit, CNOT (\CC) and SWAP (\SSS) gates.
The rest of this paper is organized as follows. In Section \ref{sec:2dAdder}, the method in \cite{Choi:2012} is discussed. We introduce the reduction techniques in Section \ref{sec:proposed}. The result of the proposed reductions is analyzed in Section \ref{sec:depth} and Section \ref{sec:qec}. We finally conclude the paper in Section \ref{sec:conc}.

\section{Quantum Addition on 2D Architectures} \label{sec:2dAdder}
In this section, we describe the circuit structure in \cite{Choi:2012} for quantum addition on 2D architectures. For an $n$-qubit quantum circuit, the method in \cite{Choi:2012} arranges the qubits in $\sqrt n \times \sqrt n$ arrays where each qubit can interact with its four neighboring qubits with no additional cost. Additionally, the circuit was divided into 3 phases which are executed sequentially. In the first phase, ripple-carry addition is performed on the first column, and carry-lookahead addition is performed on the other $\sqrt n-1$ columns. In the second phase, carry propagation is performed between columns, and finally in phase 3 carry generation and summation are performed.

In the first phase, after using a half-adder and $\sqrt n-1$ full-adders output carries $c_2, \cdots c_{\sqrt n+1}$ will be available. It is done in $32\sqrt n - 17$ unit-time steps in \cite{Choi:2012}. The carry-lookahead addition in other columns produces
\begin{eqnarray}
g_{k\sqrt{n}+j}   &   =   &   a_{k\sqrt{n}+j} \cdot b_{k\sqrt{n}+j} \label{eqn:small_g_i}   \\
p_{k\sqrt{n}+j}   &   =   &   a_{k\sqrt{n}+j} \oplus b_{k\sqrt{n}+j} \label{eqn:small_p_i}
\end{eqnarray}
for $1 \leq k \leq \sqrt n-1$ and $1\leq j \leq \sqrt{n}$. After computing $g_i$ and $p_i$ values in all columns in parallel, $G[i,j]$ and $P[i,j]$ are computed in serial based on (\ref{eqn:large_g_i}) and (\ref{eqn:large_p_i}) for $1 \leq k \leq \sqrt n-1$, and $2 \leq j \leq \sqrt n$ where $G[k\sqrt{n}+1, k\sqrt{n}+1]=g_{k\sqrt{n}+1}$ and $P[k\sqrt{n}+1, k\sqrt{n}+1]=p_{k\sqrt{n}+1}$. This part takes $34 \sqrt n - 19$ time steps in \cite{Choi:2012}. Accordingly, the first phase in \cite{Choi:2012} results in $34 \sqrt n - 19$ time steps.

\begin{eqnarray}
G[k\sqrt{n}+1,k\sqrt{n}+j] & =  & g_{k\sqrt{n}+j} \oplus \label{eqn:large_g_i} p_{k\sqrt{n}+j} \cdot G[k\sqrt{n}+1,k\sqrt{n}+j-1] \\
P[k\sqrt{n}+1,k\sqrt{n}+j] & =  & p_{k\sqrt{n}+j} \cdot P[k\sqrt{n}+1,k\sqrt{n}+j-1] \label{eqn:large_p_i}
\end{eqnarray}

In the second phase, column-level carries are computed as shown in (\ref{eqn:col_carry}) for $1 \leq k \leq \sqrt n-1$ in $18 \sqrt n - 18$ time steps.
\begin{equation}
c_{(k+1)\sqrt{n}+1}=G[k\sqrt{n}+1,(k+1)\sqrt{n}] \oplus c_{k\sqrt{n}+1} \cdot P[k\sqrt{n}+1,(k+1)\sqrt{n}] \label{eqn:col_carry}
\end{equation}

In phase 3 output carries are calculated sequentially as (\ref{eqn:carry_with_G_P}) for $1 \leq k \leq \sqrt n-1$ and $j=\sqrt n-1, ..., 1$.

\begin{eqnarray}
c_{k\sqrt{n}+j+1} =  &   G[k\sqrt{n}+1,k\sqrt{n}+j]   \oplus c_{k\sqrt{n}+1} \cdot P[k\sqrt{n}+1,k\sqrt{n}+j] \label{eqn:carry_with_G_P}
\end{eqnarray}

Finally, addition outputs are calculated as shown in (\ref{eqn:sum}) for $1 \leq k \leq \sqrt n-1$ and $1 \leq j \leq \sqrt n$. Altogether, operations in phase 3 can be performed in $18 \sqrt n+ 1$ time steps.
\begin{equation}
s_{k\sqrt{n}+j} = a_{k\sqrt{n}+j} \oplus b_{k\sqrt{n}+j} \oplus c_{k\sqrt{n}+j} \label{eqn:sum}
\end{equation}

Considering the three subcircuits for phase 1, phase 2, and phase 3 in sequence leads to $70 \sqrt n - 36$ time steps in \cite{Choi:2012}. Applying the inverse circuit to clear ancillae leads to $140
\sqrt n - 72$ time steps for the complete adder.

Based on the equations (\ref{eqn:small_g_i})-(\ref{eqn:sum}), Table \ref{table:depth-analysis} reports circuit depth in different blocks. In this table, we used the same notation in \cite{Choi:2012} for circuit blocks --- g,p to compute $g_i$, $p_i$ values in (\ref{eqn:small_g_i}) and (\ref{eqn:small_p_i}); G,P to compute $G[i,j]$ and $P[i,j]$ values in (\ref{eqn:large_g_i}) and (\ref{eqn:large_p_i}); Column\_carry to compute column-level carries in (\ref{eqn:col_carry}); Carry \& Carry1 to compute carries in (\ref{eqn:carry_with_G_P}); and SUM, SUM1 \& SUM2 to compute final outputs in (\ref{eqn:sum}).

\begin{table}[t]
\scriptsize
\caption{\label{table:depth-analysis} Basic blocks in 2D adder \cite{Choi:2012} and their depths in terms of unit-cost gates. The last term (i.e., 3) in total depth represents 2 NOTs and one CNOT gate used to construct the final output in \cite{Choi:2012}.}
\centerline{
\begin{tabular}{|l|l| p{2.5cm}|}  \hline
Name &  \#steps: gate sequence           & Circuit      \\ \hline   \hline
  {H, T, CNOT (\C), SWAP (\SSS)}             &   1 & \\
  {Toffoli (\T(a,b,0))}            & 14: 2 \SSS + 12 1-qubit & H(0)\C(b,0)T$^\dagger$(0)\SSS(b,0)\C(a,b)T(b)\C(0,b) T$^\dagger$(b)\C(a,b)\SSS(b,0)T(b)T(0)\C(a,b)H(0) T(a)T$^\dagger$(b)\C(a,b) \\
  {Half-adder(a,b,0)}       & 15: 1 \T + 1 \CC  & \T(a,b,0)\T(a,b) \\
  {Full-adder(c,a,b,0)}       & 32: 2 \T + 2 \CC + 2 \SSS & \T(a,b,0)\T(a,b)\SSS(c,a)\T(a,b,0)\T(a,b) \SSS(c,a)\\
  g,p(a,b,0)       & 15: 1 \T + 1 \CC & \T(a,b,0)\T(a,b)\\
  G,P(P,G,a,p,g,0)       & 34: 2 \T + 6 \SSS & \SSS(G,a)\SSS(P,G)T(a,p,g)\SSS(G,a)\SSS(g,0) T(a,p,g)\SSS(G,a)\SSS(P,G)\SSS(G,a)\\
  Column\_carry(P,G,C)    & 18: 1 \T + 4 \SSS  & \SSS(P,G)\T(C,G,P)\SSS(G,C)\SSS(P,G)\SSS(G,C)\\
  Carry(P,G,a,p,C)            & 18: 1 \T + 4 \SSS  & \SSS(P,G)\SSS(p,C)\SSS(a,p)\T(a,G,P)\SSS(G,a) \SSS(P,G)\\
  Carry1(p,g,c)          & 16: 1 \T + 2 \SSS  & \SSS(g,c)\T(p,g,c)\SSS(p,g)\\
  SUM(c,P,a,p)                   & 5 : 1 \CC  + 4 \SSS &\SSS(c,P)\SSS(P,a)\T(a,p)\SSS(P,a)\SSS(c,P) \\
  SUM1(c,a,p)            & 3 : 1 \CC  + 2 \SSS  & \SSS(c,a)\T(a,p)\SSS(c,a)\\
  SUM2(p,c)            & 1 : 1 \CC   & \T(c,p)\\
  \hline
  \hline
    phase 1 &   \multicolumn{2}{l|}{$34 \sqrt n - 19$: g,p + $(\sqrt n-1)$G,P} \\
    phase 2 &   \multicolumn{2}{l|}{$18 \sqrt n - 18$: $(\sqrt n - 1)$ Column\_carry}\\
    phase 3 &   \multicolumn{2}{l|}{$18 \sqrt n+ 1$: $(\sqrt n - 1)$ Carry + Carry1 + SUM1}\\
  \hline
  clearing ancillae & \multicolumn{2}{l|}{$70 \sqrt n - 39$: phase 1 + phase 2 + phase 3 - SUM1}\\
  \hline
  total depth &\multicolumn{2}{l|}{ $140\sqrt n - 72$: phase 1 + phase 2 + phase 3 + clearing ancillae + 3}\\
  \hline
  \end{tabular}}
\end{table}

\section{The Proposed 2D Adder} \label{sec:proposed}
In this section, we revise the basic blocks in \cite{Choi:2012} and introduce additional parallelism in various parts to reduce circuit depth. Basically, the proposed optimizations are based on (1) new circuit structures for CARRY and SUM basic blocks (2) reducing communication overhead in Column\_carry, (3) parallel execution of expensive Toffoli gates in G,P blocks as well as in Full-adders, and (4) reducing interaction overhead by adding concurrency to consecutive blocks.

\subsection{New Circuits} \label{sec:newcirc}
Working with the same circuit structures in \cite{Choi:2012} for Half-adder, g,p, and G,P blocks as reported in Table \ref{table:depth-analysis}, we define several new structures for the other blocks.

\begin{figure}[tb]
\centering
\scalebox{0.9}{
\input{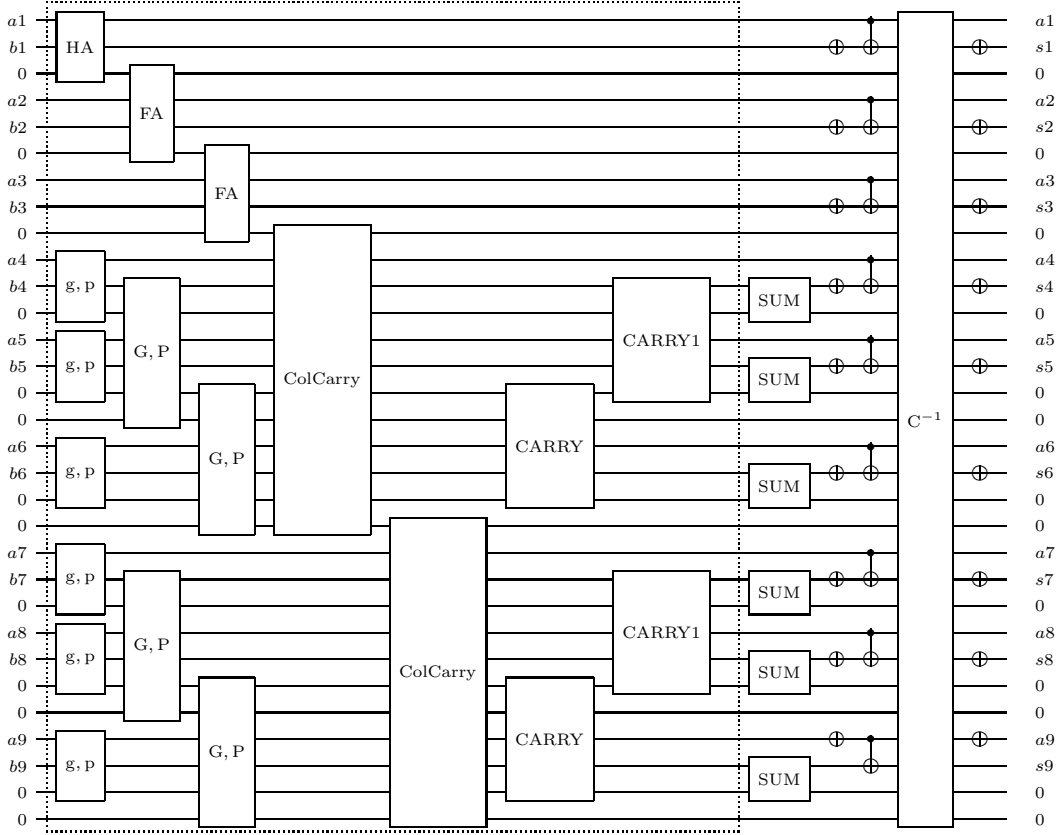}
}
\caption{ \label{fig:Adder_structure} The revised block diagram of a 2D 9-bit adder in \cite{Choi:2012} based on the blocks used in this paper. The critical path in this circuit is g,p$\dashrightarrow$G,P$\dashrightarrow$ColCarry$\dashrightarrow$ColCarry$\dashrightarrow$CARRY$\dashrightarrow$CARRY1$\dashrightarrow$SUM. The $\rm{C}^{-1}$ block is the reverse of the circuit shown in the dashed box. This reverse circuit with the NOTs and CNOTs shown are applied to clear ancillae in \cite{Choi:2012}. Except for ColCarry (Column\_carry), the number of inputs and outputs for other modules are the same as the ones shown in this figure. In Column\_carry, the number of inputs/outputs is 3 --- i.e., the first line and the last two lines are actual inputs and outputs. Note that these three lines are neighbor in the 2D layout. The qubit placement for this 2D grid and their values during the computation (up to clearing ancillae) are given in Figure \ref{fig:qubitvalues}.
}
\end{figure}

\begin{figure}[tb]
\centering
\scalebox{0.8}{
\input{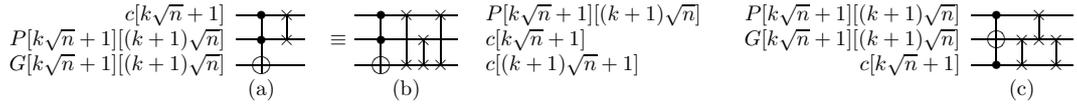}
}
\caption{(a) Circuit structure for Column\_carry based on (\ref{eqn:col_carry}). Note that $c[(k-1)\sqrt n +1]$ and $P[(k-1)\sqrt n +1][k\sqrt n]$ are not adjacent (see Figure \ref{fig:Adder_structure}). (b) Circuit in (a) with adjacent gates. (c) Circuit in (b) with relabelled qubits to show adjacent qubits.
}
\label{Fig:colcarry}
\end{figure}

\begin{figure}[tb]
\centering
\scalebox{0.8}{
\input{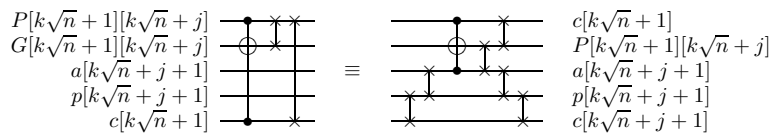}
}
\caption{Circuit structure for Carry based on (\ref{eqn:carry_with_G_P}). Inputs $a[k\sqrt n +j+1]$ and $p[k\sqrt n +j+1]$ are not used in the computation.
}
\label{Fig:Carry}
\end{figure}

\begin{itemize}
\item [$\bullet$] \textbf{Full-adder}: The first \T~and \CC ~gates in the Full-adder blocks in \cite{Choi:2012} can be executed in parallel with the gates in the Half-adder circuit. This saves one \T~and one \CC~for all $\sqrt n-1$ Full-adders.
\item [$\bullet$] \textbf{Column\_Carry}: Figure \ref{Fig:colcarry} shows the new structure of Column\_Carry block. In this circuit, $c[k\sqrt n+1]$ is from the previous column (e.g., $c_4$ in Figure \ref{fig:Adder_structure}). After the computation, the new carry, e.g., $c_7$, is moved down, to be used by the next Column\_Carry block. The previous carry, e.g., $c_4$ is placed near to the Carry module. This new structure saves 1 SWAP gate.
\item [$\bullet$] \textbf{Carry}: Figure \ref{Fig:Carry} shows the new structure for Carry block. Since $c[k\sqrt n +1]$ is required to compute all carries in different rows, $c[k\sqrt n +1]$ is moved up in this figure. On the other hand, the generated carry is required to compute sum values, and hence is moved down. This new circuit uses 5 SWAP gates (vs. 4 in \cite{Choi:2012}).
\item [$\bullet$] \textbf{SUM}: Applying the proposed circuit for Carry results in adjacent $c[k\sqrt n +j+1]$ and $p[k\sqrt n +j+1]$ values (see Figure \ref{Fig:Carry}). Based on (\ref{eqn:sum}) sum outputs can be computed by a single CNOT gate. This saves 4 SWAP gates in \cite{Choi:2012}. In order to construct $s_i$ values on $b_i$ qubits, one needs to add one SWAP gate \SSS$(p[k\sqrt n +1],c[k\sqrt n +1])$. However, this SWAP gate can be removed because of an identical SWAP gate in the Carry circuit. Accordingly, we define another circuit block Carry1 with excluding the SWAP on $c[k\sqrt n +1]$ and $P[k\sqrt n +1][k\sqrt n +j]$ (for $j=1$) qubits. We do not need to use SUM1 and SUM2 blocks in the proposed 2D adder structure.
\end{itemize}

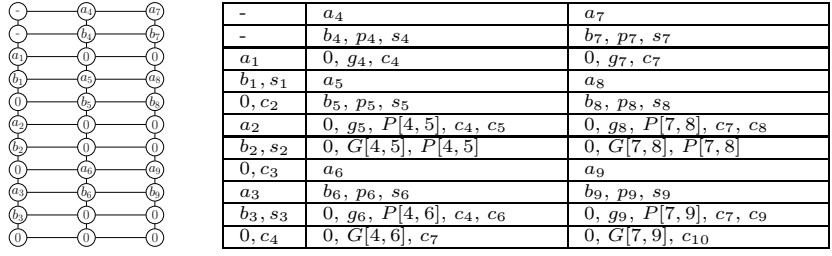
\begin{figure}[tb]
\centering
\begin{tabular}{cc}
\scalebox{0.3}{
\imagetop{\scriptsize
\begin{tikzpicture} [scale=1][font=\LARGE]

\tikzstyle{every node}=[draw,shape=circle,inner sep=0pt,minimum size=22pt];

\path (0,0)   node[label=below right:] (n00) {0};      
\path (0,1)   node[label=below right:] (n01) {$b_3$};  
\path (0,2)   node[label=below right:] (n02) {$a_3$};  
\path (0,3)   node[label=below right:] (n03) {0};      
\path (0,4)   node[label=below right:] (n04) {$b_2$};  
\path (0,5)   node[label=below right:] (n05) {$a_2$};  
\path (0,6)   node[label=below right:] (n06) {0};      
\path (0,7)   node[label=below right:] (n07) {$b_1$};  
\path (0,8)   node[label=below right:] (n08) {$a_1$};  
\path (0,9)   node[label=below right:] (n09)  {-};
\path (0,10)  node[label=below right:] (n10)  {-};

\path (3,0)   node[label=below right:] (n11) {0};
\path (3,1)   node[label=below right:] (n12) {0};    
\path (3,2)   node[label=below right:] (n13) {$b_6$};
\path (3,3)   node[label=below right:] (n14) {$a_6$};
\path (3,4)   node[label=below right:] (n15) {0};
\path (3,5)   node[label=below right:] (n16) {0};    
\path (3,6)   node[label=below right:] (n17) {$b_5$};
\path (3,7)   node[label=below right:] (n18) {$a_5$};
\path (3,8)   node[label=below right:] (n19) {0};    
\path (3,9)   node[label=below right:] (n20) {$b_4$};
\path (3,10)  node[label=below right:] (n21) {$a_4$};

\path (6,0)   node[label=below right:] (n22) {0};
\path (6,1)   node[label=below right:] (n23) {0};    
\path (6,2)   node[label=below right:] (n24) {$b_9$};
\path (6,3)   node[label=below right:] (n25) {$a_9$};
\path (6,4)   node[label=below right:] (n26) {0};
\path (6,5)   node[label=below right:] (n27) {0};    
\path (6,6)   node[label=below right:] (n28) {$b_8$};
\path (6,7)   node[label=below right:] (n29) {$a_8$};
\path (6,8)   node[label=below right:] (n30) {0};    
\path (6,9)   node[label=below right:] (n31) {$b_7$};
\path (6,10)  node[label=below right:] (n32) {$a_7$};

\draw[color=black]
(n00) -- (n11) -- (n22)
(n01) -- (n12) -- (n23)
(n02) -- (n13) -- (n24)
(n03) -- (n14) -- (n25)
(n04) -- (n15) -- (n26)
(n05) -- (n16) -- (n27)
(n06) -- (n17) -- (n28)
(n07) -- (n18) -- (n29)
(n08) -- (n19) -- (n30)
(n09) -- (n20) -- (n31)
(n10) -- (n21) -- (n32)

(n00) -- (n01) -- (n02) -- (n03) -- (n04) -- (n05) -- (n06) -- (n07) -- (n08) -- (n09) -- (n10)

(n11) -- (n12) -- (n13) -- (n14) -- (n15) -- (n16) -- (n17) -- (n18) -- (n19) -- (n20) -- (n21)

(n22) -- (n23) -- (n24) -- (n25) -- (n26) -- (n27) -- (n28) -- (n29) -- (n30) -- (n31) -- (n32)
;

\end{tikzpicture}}
}
&
\imagetop{\scalebox{1}{
\scriptsize
    \begin{tabular}{|l|p{3cm}|p{3cm}|}
        \hline
        -		     & $a_4$                  & $a_7$                  \\
        \hline
        -		     & $b_4$, $p_4$, $s_4$          & $b_7$, $p_7$, $s_7$          \\
        \hline
        $a_1$     	 & $0$, $g_4$, $c_4$            & $0$, $g_7$, $c_7$            \\
        \hline
        $b_1,s_1$ 	 & $a_5$                  & $a_8$                  \\
        \hline
        $0,c_2$   	 & $b_5$, $p_5$, $s_5$          & $b_8$, $p_8$, $s_8$          \\
        \hline
        $a_2$     	 & $0$, $g_5$, $P[4,5]$, $c_4$, $c_5$ & $0$, $g_8$, $P[7,8]$, $c_7$, $c_8$ \\
        \hline
        $b_2,s_2$ 	 & $0$, $G[4,5]$, $P[4,5]$      & $0$, $G[7,8]$, $P[7,8]$      \\
        \hline
        $0,c_3$   	 & $a_6$                  & $a_9$                  \\
        \hline
        $a_3$     	 & $b_6$, $p_6$, $s_6$          & $b_9$, $p_9$, $s_9$          \\
        \hline
        $b_3,s_3$ 	 & $0$, $g_6$, $P[4,6]$, $c_4$, $c_6$ & $0$, $g_9$, $P[7,9]$, $c_7$, $c_9$ \\
        \hline
        $0,c_4$   	 & $0$, $G[4,6]$, $c_7$          & $0$, $G[7,9]$, $c_{10}$         \\
        \hline
    \end{tabular}
}
}
\end{tabular}
\centering
\vspace{1mm}
\caption{
 \label{fig:qubitvalues} The qubit placement for the 2D grid in Figure \ref{fig:Adder_structure} and their values during the computation.
}
\end{figure}

\subsection{Reducing Communication Overhead} \label{sec:overhead}
To use adjacent gates in the 2D quantum adder, we use a set of SWAP gates inside each circuit block. The added SWAP gates are used for communication between those gates required for the computation. In other words, the added SWAP gates are not required for the computation, and should be reduced as much as possible. Independent optimization of different blocks can reduce communication overhead inside each subcircuit, but has no view about the neighboring subcircuits. In this section, we consider consecutive circuit blocks to reduce communication overhead further. Note that the optimizations given in this section are based on the new circuit blocks given in Section \ref{sec:newcirc}.

\begin{figure}[tb]
\centering
\scalebox{0.8}{
\input{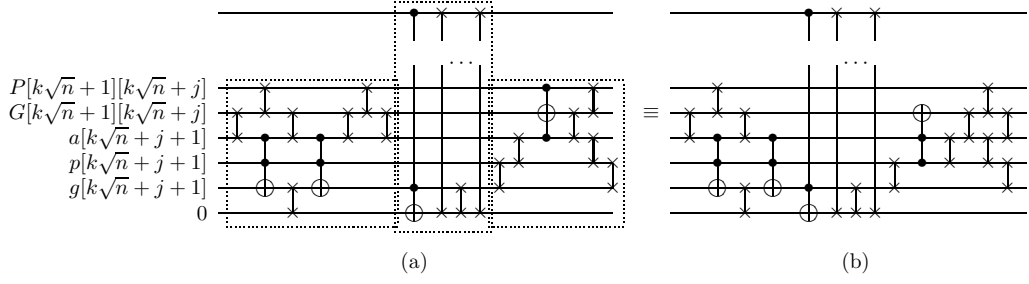}
}
\caption{(a) G,P, Column\_carry, and Carry blocks in cascade. The three rightmost SWAP gates in G,P can be merged with gates in the Carry block to construct a new circuit shown in (b).
}
\label{Fig:GPCarry}
\end{figure}

\begin{figure}[tb]
\centering
\scalebox{0.8}{
\input{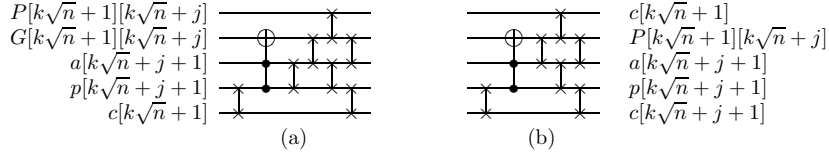}
}
\caption{New circuit structures for Carry (a) and Carry1 (b) based on the optimization shown in Fig \ref{Fig:GPCarry1}. Note that the first SWAP gate can be executed in parallel with gates in the previous block (see Figure \ref{Fig:GPCarry1}).
}
\label{Fig:GPCarry1}
\end{figure}

\begin{figure}[t]
\centering
\scalebox{0.8}{
\input{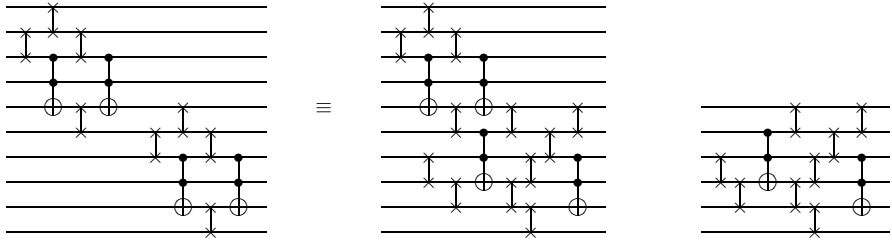}
}
\caption{Construction of $G[k\sqrt{n}+1,k\sqrt{n}+j+1]$ can be done in parallel to construction of $P[k\sqrt{n}+1,k\sqrt{n}+j-1]$ in two consecutive G,P blocks. The right circuit shows the new circuit structure for G,P (except for the first G,P block).
}
\label{Fig:GPGP}
\end{figure}

\begin{itemize}
\item [$\bullet$] \textbf{G,P $\Rrightarrow$ Carry}: Reconsider (\ref{eqn:large_g_i}), (\ref{eqn:large_p_i}), and (\ref{eqn:carry_with_G_P}) and note that the result of Column\_carry in (\ref{eqn:col_carry}), i.e., $c[k\sqrt n +1]$, is constructed on the last qubit in the Carry block (see Figure \ref{Fig:colcarry} and Figure \ref{Fig:Carry}). Figure \ref{Fig:GPCarry} shows the blocks in sequence. To simplify the circuit, note that the last three SWAP gates in G,P can be moved to right. Next, the resulting circuit can be reconstructed as shown in Figure \ref{Fig:GPCarry}(b). Accordingly, three SWAP gates in each G,P block can be saved. Figure \ref{Fig:GPCarry1} shows the new circuits for Carry and Carry1. Note that some of G,P blocks are directly connected to the Carry (or Carry1) blocks without any interaction with Column\_carry blocks. For such cases, we can apply the same mechanism.
\item [$\bullet$] \textbf{G,P $\Rrightarrow$ G,P}: Each G,P block constructs two outputs based on (\ref{eqn:large_p_i}) and (\ref{eqn:large_g_i}) where $G[k\sqrt{n}+1,k\sqrt{n}+j]$ depends on $G[k\sqrt{n}+1,k\sqrt{n}+j-1]$ and $P[k\sqrt{n}+1,k\sqrt{n}+j]$ depends on $P[k\sqrt{n}+1,k\sqrt{n}+j-1]$. Since $G[k\sqrt{n}+1,k\sqrt{n}+j]$ is constructed first, we can use it to construct $G[k\sqrt{n}+1,k\sqrt{n}+j+1]$ in parallel to construction of $P[k\sqrt{n}+1,k\sqrt{n}+j-1]$. This can save one Toffoli and one SWAP. Figure \ref{Fig:GPGP} shows the result of this optimization.
\end{itemize}

\section{Depth Analysis} \label{sec:depth}
In this section, we analyze the circuit depth of a 2D $n$-bit quantum adder based on the circuit structures proposed for each block.

\begin{itemize}
\item [$\bullet$] \textbf{Phase 1 --- Half-adder+Full-adder}: We can execute Half-adder and the first two gates (\T+\CC) in all Full-adders in parallel. This results in 1\T+1\CC+($\sqrt n -1$)(2\SSS+1\CC+1\T) time steps.
\item [$\bullet$] \textbf{Phase 1 --- g,p+G,P}: Each g,p block includes one Toffoli gate and one CNOT gate. Except for the first G,P block, the other $\sqrt n -2$ G,P blocks include 3 SWAPs and 1 Toffoli. The first G,P block includes two Toffoli and two SWAP gates. Altogether, circuit depth can be calculated as (1\T+1\CC)+(2\T+2\SSS)+$(\sqrt n-2)$(3\SSS+1\T).
\item [$\bullet$] \textbf{Phase 2 --- Column\_carry}: There are $\sqrt n-1$ Column\_carry blocks in cascade. This results in $\sqrt n-1$(1\T+3\SSS) time steps.
\item [$\bullet$] \textbf{Phase 3 --- Carry + SUM}: There are $\sqrt n - 2$ Carry blocks followed by one Carry1 block and one SUM block. Therefore, circuit depth is $(\sqrt n-2)$(1\T+4\SSS)+(3\SSS+1\T)+1\CC.
\end{itemize}

Table \ref{table:ourdepth} reports circuit depth for each component and the total depth in the proposed 2D quantum adder. As can be seen in this table, circuit depth is improved by a factor of $\frac{26}{35}$ (i.e., \%24).

\begin{table}
\scriptsize
\caption{\label{table:ourdepth} Circuit depth for our blocks in 2D adder. Circuit depths for CNOT (\CC), SWAP (\SSS), and Toffoli (\T) gates are considered as 1, 1, and 14 as done in \cite{Choi:2012}. }
\centerline{
    \begin{tabular}{|l|l|l|l|l|l|}
        \hline
        Block & Circuit & Ours & \cite{Choi:2012} \\
        \hline
        Half-adder        & 1\T+1\CC                                         & 15               & 15\\ Full-adder        & 2\SSS+1\CC+1\T                                   & 17               & 32\\
        g,p               & 1\T+1\CC                                         & 15               & 15\\
        G,P (first)       & 2\T+2\SSS                                        & 30               & 34\\
        G,P (others)      & 3\SSS+1\T                                        & 17               & 34\\
        Column\_carry     & 1\T+3\SSS                                        & 17               & 18\\
        Carry             & 1\T+4\SSS                                        & 18               & 18\\
        Carry1            & 3\SSS+1\T                                        & 17               & 18\\
        SUM               & 1\CC                                             & 1                & 5\\
        \hline
        Phase1-1          & 1\T+1\CC+($\sqrt n -1$)(2\SSS+1\CC+1\T)          & 17$\sqrt n -2$   & 32$\sqrt n -17$\\
        Phase1-2          & (1\T+1\CC)+(2(\T+2\SSS)+$(\sqrt n-2)$(3\SSS+1\T) & 17$\sqrt n + 11$ & 34$\sqrt n -19$\\
        Phase2            & $(\sqrt n-1)$(1\T+3\SSS)                           & 17$\sqrt n -17$  & 18$\sqrt n -18$\\
        Phase3            & $(\sqrt n-2)$(1\T+4\SSS)+(3\SSS+1\T)+1\CC        & 18$\sqrt n-18$   & 18$\sqrt n +1$\\
        \hline
        clearing ancillae & Phase1-2+Phase2+Phase3-SUM                       & 52$\sqrt n-24$   & 70$\sqrt n -39$\\
        \hline
        2D Adder          & Phase1-2+Phase2+Phase3+clearing ancillae+3                                                & 104$\sqrt n-46$  & 140$\sqrt n -72$\\
        \hline
    \end{tabular}}
\end{table}

\begin{figure}[tb]
\scriptsize
\scalebox{1}{
\input{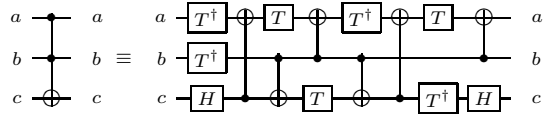}
}
\centering
\vspace{-1mm}
\caption{
 \label{fig:Toffoli2} Toffoli decomposition with depth 6\C+2  \cite{Amy12}.
}
\vspace{-1mm}
\end{figure}

\begin{figure}[tb]
\scriptsize
\scalebox{1}{
\input{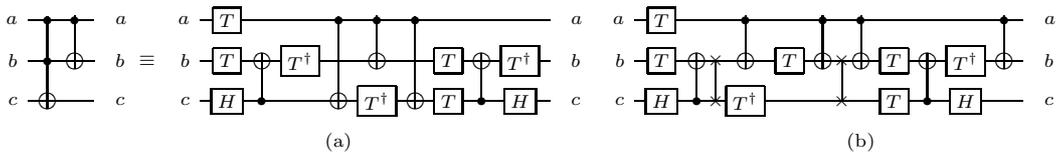}
}
\centering
\vspace{-1mm}
\caption{
 \label{fig:Peres} (a) Peres decompositions with depth 5\C+3 \cite{Amy12}, (b) Toffoli with adjacent gates based on Peres decomposition (depth=6\C+2\SSS+4).
}
\vspace{-1mm}
\end{figure}

In \cite{Amy12}, a new circuit for Peres with depth=5\C+3 has been proposed (Figure \ref{fig:Peres}(a)). After inserting one CNOT (to have Toffoli) and two SWAP gates to have adjacent gates, one can use the new circuit with depth=6\C+2\SSS+4 in order to further optimize the proposed 2D adder. Note that in \cite{Amy12}, a circuit structure for Toffoli gate with depth=6\C+2 has been proposed too, Figure \ref{fig:Toffoli2}. However, working with Peres gate results in a more compact circuit in terms of the number of SWAP gates. Following this path results in depth=92$\sqrt n$+const for the proposed 2D quantum adder. Table \ref{table:ourdepth2} compares circuit depth based on different costs for Toffoli and SWAP gates.

\begin{table}
\scriptsize
\caption{\label{table:ourdepth2} Circuit depth for the proposed adder and the one in \cite{Choi:2012} considering different costs for Toffoli and SWAP gates. }
\centerline{
    \begin{tabular}{|c|c|c|c|c|c|c|c|}
        \hline
         \multicolumn{2}{|l|}{\T-depth=14,\SSS-depth=1} &          \multicolumn{2}{l|}{\T-depth=14,\SSS-depth=3} &
         \multicolumn{2}{l|}{\T-depth=12,\SSS-depth=3} &
         \multicolumn{2}{l|}{\T-depth=12,\SSS-depth=1} \\
        \hline
        Ours         & \cite{Choi:2012}       & Ours         & \cite{Choi:2012}             & Ours         & \cite{Choi:2012}            & Ours         & \cite{Choi:2012}             \\ \hline
        \hline
        104$\sqrt n$ & 140$\sqrt n $ & $144\sqrt n$  & $176\sqrt n$  & $132\sqrt n$  & 160$\sqrt n $ & 92$\sqrt n $ & 124$\sqrt n $ \\
        \hline
    \end{tabular}
    }
\end{table}

\section{Error Correction}  \label{sec:qec}
To protect quantum information from errors due to e.g., noise or decoherence, quantum error correction (QEC) should be used in any large-scale quantum computation. In the recent years, various models for {QEC} have been proposed \cite{Nielsen00}. A common technique, known as concatenated quantum code, is to encode a logical qubit into the state of several physical qubits (e.g., 7 in Steane code and 9 in Bacon-Shor code \cite{Nielsen00}, both for one level of concatenation).

Let assume each unitary operation should be followed by quantum error correction for proper computation. This results in an aggressive quantum error correction mechanism. In some circumstances, one may insert error correction after several operations, instead of each operation.
Consider a quantum computation $U$ with $N_U$ logical operations which include only FT quantum gates. Moreover, assume that error correction for each FT gate requires $N_E$ physical instructions. $N_E$ includes SWAPs required for communication. Normally, $N_E$ differs for various logical operations; however, we can consider the worst-case value among all FT gates. Working with concatenated quantum error correction techniques, the total physical gate count at concatenation level $L$ can be estimated as
$N_L = N_{L-1}+N_{L-1} \times N_E$ or $N_L \approx N_{L-1} \times N_E$. We have $N_0=N_U$, and therefore, $N_L =N_U (N_E)^L$. Accordingly, besides the effect of the proposed approach on circuit depth, one can implement the proposed 2D adder with fewer gates --- the reduction factor is $\frac{24}{35}$.

\section{Conclusion} \label{sec:conc}
We considered a quantum adder on 2D quantum architectures. Our work is based on the results reported in \cite{Choi:2012} with several improvements. In particular, we optimized the building blocks of the 2D adder with focus on reducing the communication overhead required in 2D quantum architectures. Having optimized consecutive blocks, the proposed adder can execute expensive Toffoli gates concurrently in several locations. The suggested optimizations improve depth=$140\sqrt n+k_1$ in \cite{Choi:2012} to $92\sqrt n+k_2$ for constants $k_1$ and $k_2$.


\section*{Acknowledgements}
Authors were supported by the Intelligence Advanced
Research Projects Activity (IARPA) via Department
of Interior National Business Center contract number
D11PC20165. The U.S. Government is authorized to reproduce
and distribute reprints for Governmental purposes
notwithstanding any copyright annotation thereon. The views
and conclusions contained herein are those of the authors and
should not be interpreted as necessarily representing the official
policies or endorsements, either expressed or implied, of
IARPA, DoI/NBC, or the U.S. Government.

\begin{landscape}
\begin{figure}[tb]
\scriptsize
\scalebox{1}{
\input{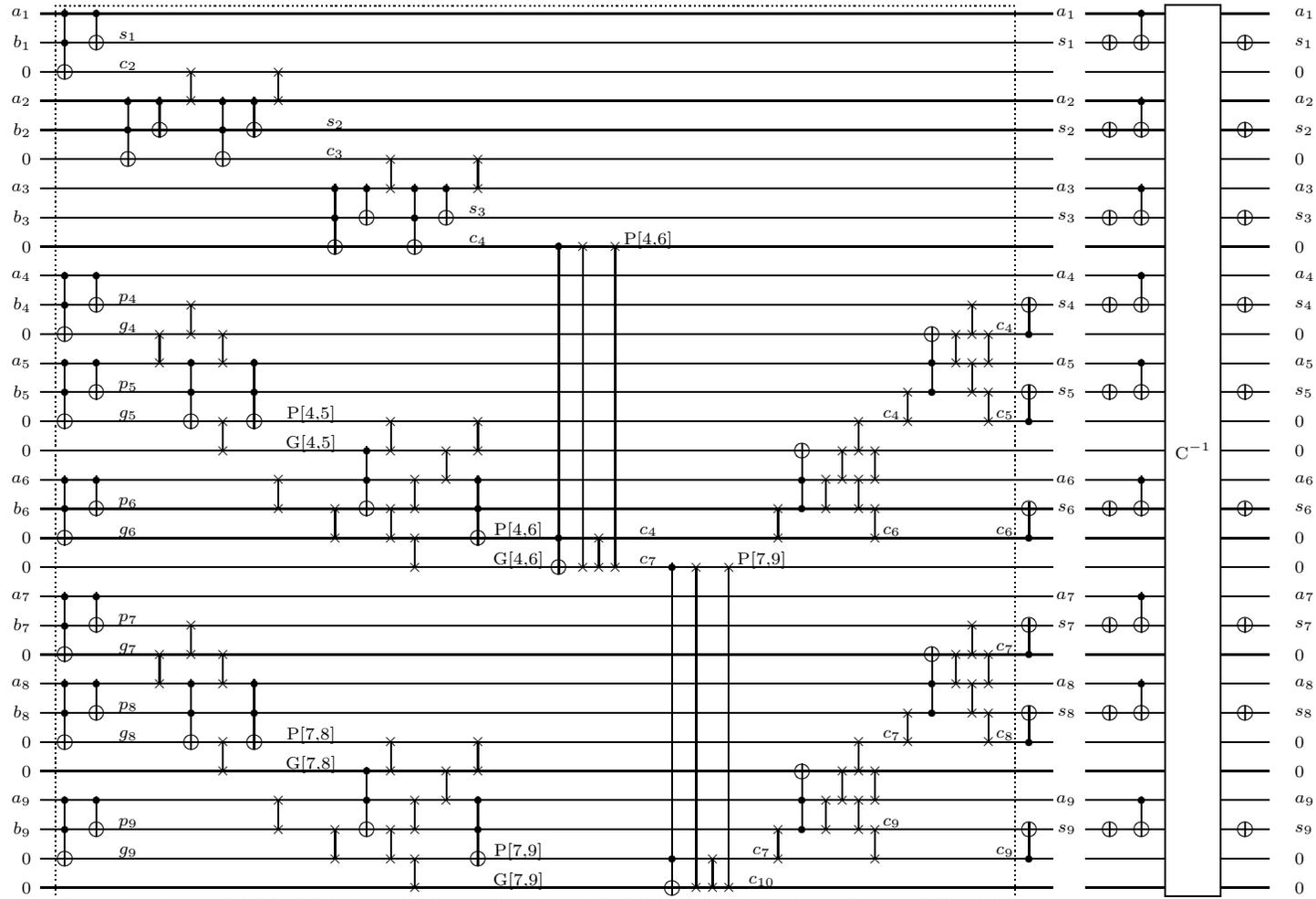}
}
\centering
\vspace{-1mm}
\caption{
 \label{fig:CompleteAdder} A 9-bit adder based on the proposed blocks. Carry, $G_{i,j}$, $p_i$, and $g_i$ values are shown in this figure. The $\rm{C}^{-1}$ block is the reverse of the circuit shown in the dashed box applied with the NOTs and CNOTs shown to clear ancillae. All gates use adjacent gates in the 2D layout. For qubit locations see the table in Figure \ref{fig:qubitvalues}.
}
\vspace{-1mm}
\end{figure}
\end{landscape}
\end{document}